\documentclass[12pt]{article}
\pdfoutput=1
\usepackage{cite}
\usepackage{amssymb}
\usepackage{amsmath}
\usepackage{bm} 
\usepackage{graphicx}
\usepackage{color}
\usepackage{xcolor}
\usepackage{dsfont}
\usepackage{enumerate}
\usepackage{hyperref}
\usepackage{setspace}
\usepackage[latin1]{inputenc}
\definecolor{nicered}{rgb}{0.7,0.1,0.1}
\definecolor{nicegreen}{rgb}{0.1,0.5,0.1}
\hypersetup{colorlinks,citecolor= nicegreen,linkcolor= nicered}

\newcommand{\be}  {\begin{equation}}
\newcommand{\ee}  {\end{equation}}

\def\e6{E(6)}
\def\10{SO(10)}
\def\21{SA(2) $\otimes$ U(1) }
\def\321{$\mathrm{SU(3) \otimes SU(2) \otimes U(1)}$ }

\def\422{SA(4) $\otimes$ SA(2) $\otimes$ SA(2)}
\def\roughly#1{\mathrel{\raise.3ex\hbox{$#1$\kern-.75em
      \lower1ex\hbox{$\sim$}}}} \def\lsim{\roughly<}
\def\gsim{\roughly>}

\def\lsim{\raise0.3ex\hbox{$\;<$\kern-0.75em\raise-1.1ex\hbox{$\sim\;$}}}
\def\gsim{\raise0.3ex\hbox{$\;>$\kern-0.75em\raise-1.1ex\hbox{$\sim\;$}}}
\nopagebreak
\allowdisplaybreaks
\begin{document}
\begin{titlepage}

  \newcommand{\AddrLiege}{{\sl \small IFPA, Dep. AGO, Universite de
      Liege, Bat B5,\\ \small \sl Sart Tilman B-4000 Liege 1,
      Belgium}}
  \newcommand{\AddrIPM}{{\sl \small Physics school, Institute for
      Research in Fundamental Sciences (IPM),\\ \sl \small
      P.O. Box 19395-5531, Tehran, Iran}}
  \vspace*{0.5cm}
\begin{center}
  \textbf{\large Two-loop snail diagrams: relating neutrino masses}\\[3mm]
  \textbf{\large to dark matter}
  \\[9mm]
  Yasaman Farzan\footnote{e-mail address:{\tt yasaman@theory.ipm.ac.ir}}
  \vspace*{0.4cm}\\
  \AddrIPM.
  \vspace{1cm}\\
\end{center}
%
\begin{abstract}
  \onehalfspacing
  Various mechanisms have been developed to explain the origin of
  Majorana neutrino masses. One of them is  radiative mass
  generation. Two-loop mass generation is of particular interest
  because the masses and couplings of new particles propagating in the
  loop can be in the range testable by other experiments and
  observations. In order for the radiative mass suppression to be
  reliable, it should be guaranteed that lower loop contributions are
  suppressed.  Based on loop topology and the form of electroweak
  presentation of the particles propagating in the loop, one can
  determine whether a lower---and therefore dominant---loop
  contribution is possible. We present a model based on these general
  considerations which leads to neutrino masses via a two-loop diagram
  which we dub as ``snail-diagram''. The model has two natural
  candidates for dark matter one of them being a neutral Dirac fermion
  which can satisfy the conditions of the thermal freeze-out scenario
  by annihilation to lepton pairs. We comment on the possibility of
  explaining the GeV gamma ray excess observed by Fermi-LAT  from the
  region close to the Galaxy Center. We also discuss  possible signals at the LHC and at experiments searching for lepton flavor violating rare decays.
\end{abstract}
\end{titlepage}
\setcounter{footnote}{0}
\section{Introduction}
\label{sec:intro}
Origin of neutrino masses and nature of  Dark Matter (DM) are  among
the most compelling open questions in particle physics.  In recent years,
models in which neutrinos acquire their masses at  loop level have
received considerable attention (see
Ref. \cite{Bonnet:2012kz,Farzan:2012ev,Angel:2012ug,Sierra:2014rxa} for
a model-independent analysis). Within these models, the smallness of
neutrino masses can be understood (at least partially) by loop
suppression. If the new particles propagating in the loop are lighter
than a few TeV, the resulting scheme will be phenomenologically
interesting because in that case the new states can potentially be
produced at the LHC. If this turns out to be the case, the radiative neutrino mass model can be tested at man-made accelerators. This is a great advantages over
the ``canonical''
tree-level type-I seesaw model \cite{seesaw}, for which on-shell
production of the new states is inconceivable in any foreseeable
future in  man-made accelerators.

 Assuming that the only source of electroweak symmetry breaking is the vacuum expectation of the Higgs, $n$-loop contributions to neutrino masses can be estimated as
\begin{equation}
  \label{eq:order-mag}
  m_\nu\sim
  \left(\frac{g^2}{16 \pi^2}\right)^n
  \left(\frac{\langle H\rangle^2}{m_\text{New}}\right)
  \left[1,\left(\log\frac{\Lambda}{m_\text{New}}\right)^n\right]\quad
  \ ,
\end{equation}
where $m_\text{New}$ is the mass scale characterizing the new physical
degrees of freedom appearing in the loop and $\Lambda$ is the
ultraviolet (UV) cut-off scale of the model satisfying
$\Lambda\gg m_\text{New}$. Taking $m_{\text{New}}\sim 1$~TeV, $m_\nu \sim 0.1-1$~eV
\cite{Forero:2014bxa,GonzalezGarcia:2012sz,Fogli:2012ua}, $\Lambda/m_\text{New}\sim 10$ and  $n=2$, we find that $g\sim 10^{-3}$. Increasing $n$, the required values of the couplings will of course increase. The same couplings also lead to Lepton Flavor Violating (LFV) processes. For $m_\text{NEW}<10$~TeV, null results of searches for LFV rare decays of the muon and the tau lepton yield strong bounds on the combinations of such couplings.  For $n=2$, these bounds are naturally satisfied but for $n>2$, a special mechanism such as the flavor symmetries suggested in
\cite{Farzan:2012ev} have to be invoked to make neutrino masses consistent with LFV bounds. From this perspective, the two-loop neutrino mass models seem more natural and are favored over higher order loop models.

In order to explain the smallness  of neutrino masses through
radiative schemes, one should make sure that lower---and therefore
dominant---loop contributions are absent. In  \cite{Farzan:2012ev} based on general
considerations of topological structure of the loops and symmetries,
 the requirements assuring the absence of lower order contributions have been systematically formulated. Here in this
paper, using the ``recipes and ingredients'' outlined in
\cite{Farzan:2012ev}, we reconstruct a model where neutrino masses are
generated at the two-loop level through what we call ``snail
diagrams''.

Our model respects a new $Z_2\times U(1)_\text{New}$ symmetry. These
symmetries stabilize two of the lightest particles with non-trivial
transformation under these discrete symmetries against decay. If these
stable particles are neutral, they may be considered as a candidate
for DM. In our model, a Dirac fermion, $\psi$ which is a singlet under
the electroweak symmetry plays the role of the DM. The DM couples to
left-handed leptons via a Yukawa coupling. The abundance of $\psi$ is
determined by thermal freeze-out scenario via annihilation to lepton
pairs.  To avoid the severe bounds from LFV, we assume that $\psi$
couples exclusively to only one flavor. 
An excess in the GeV range $\gamma$-ray has
been reported in Fermi-LAT data on signal from regions close to
galactic center. One of the solutions is dark matter of mass 10 GeV
annihilating into tau pair \cite{Celine}. Another possibility is annihilation into $b \bar{b}$ pair \cite{McCabe}. The dark matter origin of this signal has
been however questioned and alternative sources have been suggested \cite{Gabi}. We will
comment on the possibility of accommodating this scenario within our
model.  

The paper is organized as follows. In section 2, we generally discuss two-loop contributions to neutrino masses based on the topology of the diagrams. In section 3, we introduce the content of the model. In section 4, we discuss lepton flavor violating effects. In section 5, we calculate the contribution to neutrino masses. In section 6, we discuss the annihilation of dark matter pair and possibility of accommodating the claimed gamma ray excess from the region close to the galactic center.  In sections 7 and 8, we respectively discuss signatures at the LHC and contribution to anomalous magnetic dipole moment. Conclusions are summarized in section 9.
\section{Comments on two-loop neutrino masses: crab and snail
  diagrams}
\label{sec:setup}
\begin{figure}[t!]
\centering

\includegraphics[scale=0.9]{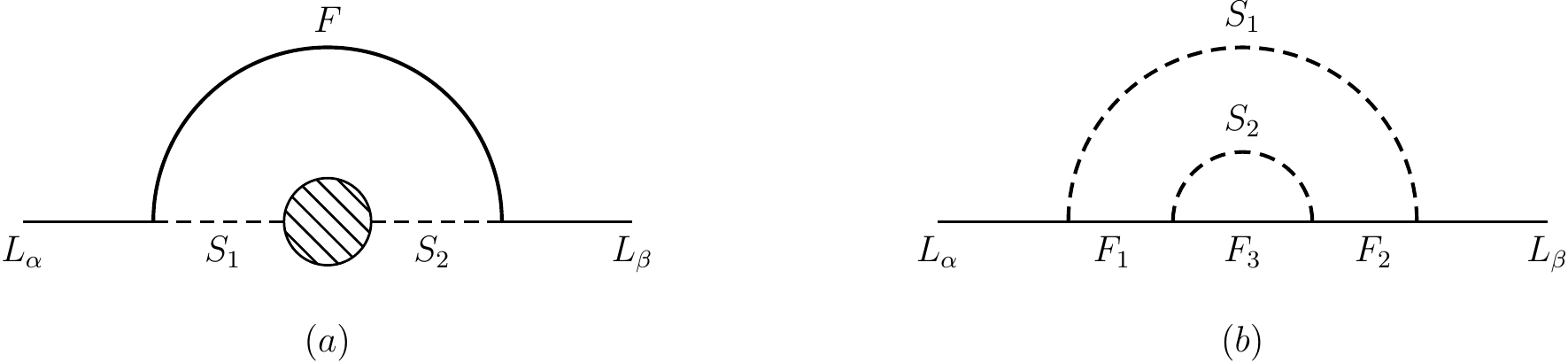}
\caption{\it Two-loop diagrams with one-loop wave function
renormalization of scalar ($a$) and fermion ($b$) fields.}
\label{fig:wave-func-diag}
\end{figure}
Two-loop diagrams contributing to neutrino masses have been
systematically discussed in \cite{Farzan:2012ev,Sierra:2014rxa}. Based on the
topologies of the two-loop diagrams, they can be classified in two
groups: $(1)$ Diagrams with a one-loop sub-diagram that can be
considered as a correction to one of the internal lines.
Figs. (\ref{fig:wave-func-diag}- $a)$  and  (\ref{fig:wave-func-diag}- $b)$ show corrections to internal scalar and fermion lines, respectively. The ``bubble'' on the scalar
line may indicate a fermion loop, a scalar loop with trilinear scalar
vertices or a scalar loop with quartic scalar vertex.
Further details can be found in \cite{Farzan:2012ev}. $(2)$ Diagrams in which an
internal line interconnects the scalar and fermion lines coming from
the vertex connected to the external lines.  These types of diagrams are
rather well-known and have been employed in the literature to
radiatively produce neutrino mass at the two-loop level. A pioneer
work using such diagram is the famous Cheng-Li-Babu-Zee model
\cite{Cheng:1980qt,Zee:1985id,Babu:1988ki}.

In Ref. \cite{Farzan:2012ev}, it is argued that diagrams of type
(\ref{fig:wave-func-diag}-$a)$ contributing to the effective Weinberg
operator
\begin{equation}
  \label{eq:weinberg-operator}
  {\cal O}_5\sim
\left(L^T\,C\,i\tau_2\,H\right)\left(H^T\,i\tau_2\,L\right)\ ,
\end{equation}
can always be accompanied by a one-loop contribution to neutrino mass.
The reason is that if the symmetries of the Lagrangian allow the
one-loop internal sub-diagram, they will also allow  a
renormalizable term with which the internal loop can be replaced.
Depending on where the two external Higgs lines are attached (vacuum
insertions $\langle H\rangle$), these renormalizable terms can be
$S_1\,S_2$, $S_1\, S_2\,H$ or $S_1\,S_2\,H^2$.

\begin{figure}[t!]
 \centering
  \includegraphics[scale=0.9]{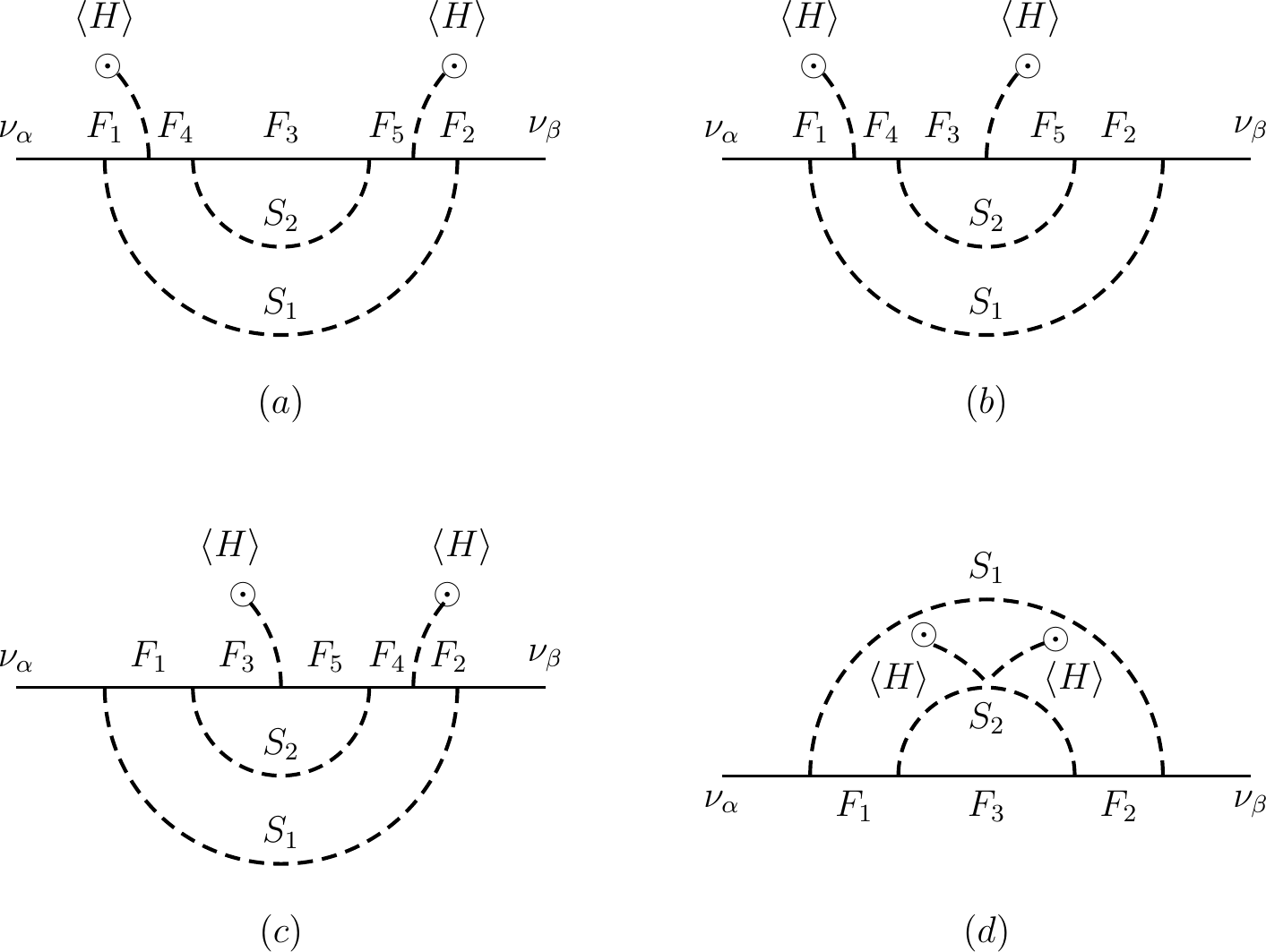}
  \caption{\it Generic crab and snail diagrams.
   We have not specified on which fermionic line the chirality flip takes place.}
  \label{fig:crab-plus-snail}
\end{figure}
On the contrary, the so-called rainbow diagrams generically depicted in Fig. (\ref{fig:wave-func-diag}-$b)$  are not necessarily accompanied by any one-loop 
counterpart.  The argument is based on the following fact. While a
term such as $S_1S_2H^2$ is renormalizable, its fermionic counterpart,
$F_1\,F_2\,H^2$, is not. Thus, depending on the electroweak structure
of the fermion lines attached to the internal loop ($F_1$ and $F_2$ in
Fig. \ref{fig:wave-func-diag}-$(b)$) and the way in which the Higgs
external lines are attached to the corresponding diagram, there might
or might not be a one-loop contribution.

For the sake of the following discussion, let us consider the diagrams
in Fig. \ref{fig:crab-plus-snail}: ``crab'' (diagrams $(a)-(c)$) and
``snail'' diagrams (diagram $(d)$). The internal loops in ``crab''
diagrams can be respectively replaced by renormalizable vertices
$F_4\,F_5$, $F_4\,F_2\,H$ and $F_1\,F_4\,H$. ``Crab'' diagrams are
therefore always accompanied by a leading one-loop contribution, and
are in that sense irrelevant. For ``snail'' diagrams, instead, there
is no such possibility because $F_1 F_2 H^2$, being non-renormalizable
cannot appear in the Lagrangian.  This argument of course holds under
the assumption that neutrino masses are generated below the
electroweak symmetry breaking scale only from Weinberg operator in Eq.
(\ref{eq:weinberg-operator}). If we included a hypercharge $-2$
 electroweak scalar triplet ($\Delta$), with scalar interactions
enabling a non-vanishing vacuum expectation value, $\langle
\Delta\rangle\neq 0$, the external Higgs lines (vacuum insertions
$\langle H\rangle$) could be replaced by a single triplet vacuum
insertion $\langle \Delta \rangle$. In that case the internal loop
could be replaced by the renormalizable vertex $\Delta\,F_1\,F_2$.

In what follows we build a model where the effective Weinberg operator
arises via a ``snail'' diagram.

\section{Snail models}
\label{sec:snail-model-DM}
In this section, we present a model that can provide a suitable Dirac
fermion DM and give mass to neutrinos via a two loop diagram. We first
introduce the symmetry structure and field content of the model and
then discuss why each assumption is made. In the next sections, we shall
discuss the contribution to neutrino mass, annihilation of DM pairs to
lepton pairs, effects on LFV and magnetic dipole moment of the
muon and signals at the LHC.

 The model is based on an unbroken $Z_2 \times U(1)_{NEW}$ symmetry. The
SM particles are all even and neutral under this symmetry. The model
also enjoys an approximate lepton number symmetry, $U(1)_L$ softly
broken by a fermion mass mixing term. The field content of the model
is shown in table \ref{tabl}.

\begin{table}[htb]
\begin{center}
\begin{tabular}{|c|c|c|c|c|c|c|}
\hline
~& $SU(2)$ &  $U(1)_Y$ &  $U(1)_{L}$ & $U(1)_{NEW}$&  $Z_2$ \\
\hline
$F_1$ & d & -1 & 1 & 1  & +   \\
$F_2$ & d&- 1 & 1 & -1  & + \\
$F_3$ & d& 1 & 1 & 1 &  + \\

$\psi$ &s & 0 & 1 & 1 &- \\
\hline
$S$ & s &0 &0 & -1 &  + \\
$\Phi$ & d & -1 & 0& 0  & - \\
$\Phi^\prime$ & d & -1 & 0 & -1 &  - \\
\hline
\end{tabular}
\caption{Field content of the model. By ``d" and ``s" in the second column we mean  doublet and singlet, respectively. We have used the convention for hypercharge in which $Q=T^3+Y/2$.  The first four fields  ({
\it i.e.,} $F_1$, $F_2$, $F_3$ and $\psi$) are Dirac fermions and  the last three lines  ($S$, $\Phi$ and $\Phi^\prime$) are scalar fields. \label{tabl}}
\end{center}
\end{table}
The new fermions are all Dirac particles and their masses are of form
$$\sum_i m_{F_i} \bar{F}_iF_i+m_\psi \bar{\psi}\psi \ .$$
As a result, neutral and charged components of $F_i$ are degenerate.
We also include mass term of form
\be \label{mM} m_M (F_{2R}^a)^TcF_{3R}^b\epsilon_{a b}+m_M^\prime (F_{2L}^a)^TcF_{3L}^b\epsilon_{a b}+ {\rm H.c.}\ee
which is supposed to be the only source of lepton number violation.
The Yukawa couplings of the new particles symmetric under $Z_2 \times U(1)_{NEW} \times U(1)_L$ are
\be \label{Lagrangian}
\mathcal{L}_{Yukawa}= g_\alpha S^\dagger F_{1R}^\dagger L_\alpha+h_\alpha S F_{2R}^\dagger L_\alpha+Y_{R\alpha}\Phi^{\prime\dagger}\psi_R^\dagger  L_\alpha+
\ee
$$ Y_1 \Phi^\dagger \psi_L^\dagger F_{1R}
+Y_2 \epsilon_{ab}\Phi^a \psi_L^\dagger F_{3R}^b + Y_1^\prime \Phi^\dagger \psi_R^\dagger F_{1L}
+Y_2^\prime \epsilon_{ab}\Phi^a  \psi_R^\dagger F_{3L}^b+ {\rm H.c.} $$

The new scalars can have interactions between themselves and SM Higgs. We assume that only the SM Higgs obtains a VEV so $U(1)_{NEW}$ and the new $Z_2$  symmetries remain unbroken. The $Z_2$ and $U(1)_{NEW}$ forbid mass terms mixing the scalars such as $H^\dagger \Phi$ or $\Phi^\dagger \Phi^\prime$. We can however have  couplings of form
$$(\lambda (H^a \Phi^b \epsilon_{ab})^2+{\rm H.c})   \ \ \ {\rm and} \ \ \ \lambda^\prime |H^\dagger \Phi|^2.$$  The $\lambda$ coupling  after electroweak symmetry breaking will lead to a mass term of form $(\Phi^0)^2$ for the neutral component
 of $\Phi^0\equiv (\phi_R+i\phi_I)/\sqrt{2}$. Thus, there will be a splitting between $\phi_R$ and $\phi_I$. We however take $\lambda$ to be real so these fields remain mass eigenstates. We will denote the masses of these components
with $m_I$ and $m_R$:
$$m_R^2-m_I^2= \lambda \langle H^0 \rangle^2.$$
 The couplings of $\phi_R$ ($\phi_I$) to $F_1$ and $F_2$ are respectively given by $Y_1/\sqrt{2}$ ($iY_1/\sqrt{2}$)
and $Y_2/\sqrt{2}$ ($iY_2/\sqrt{2}$). Notice that $U(1)_{NEW}$ protects real and imaginary components of $S$ as well as the neutral component of $\Phi^\prime$ from such splitting.
The $\lambda^\prime$ coupling leads to a mass term of form $\lambda^\prime \langle H^0\rangle^2 |\phi^-|^2$. Taking $\lambda^\prime$ positive, $\phi^-$ can be heavier than $\phi_I$ and
$\phi_R$ so $\phi^-$ can decay to $\phi_R$ and/or $\phi_I$.

Imposing both the $Z_2$  and $U(1)_{NEW}$ symmetries opens  a possibility of having two DM candidates. 
  The neutral components of $F_i$ cannot  be suitable dark matter candidates in this model because, as mentioned above, charged components of $F_i^-$
  are also degenerate with them and might lead to the presence of electrically charged DM. Thus, we take $F_i$ heavy enough to decay to $\psi$ and $\Phi$. In this case, $\phi_I$ which is the lightest
  $U(1)_{NEW}$ neutral and $Z_2$-odd particle will be stable and contribute to the dark matter abundance. If $\phi_I$ and $\phi_R$ are quasi-degenerate ({\it i.e.,} $(m_R-m_I)/m_R<1/20$), their
  contribution to DM abundance will be suppressed within thermal freeze-out scenario. The electroweak singlet $S$ can  also kinematically be made stable and can therefore contribute to DM abundance.
 The annihilations of $S$ will  be then through the $g_\alpha$ and $h_\alpha$ couplings to $l \bar{l}$ pairs. The annihilation will be suppressed by $m_l^2/m_F^2 \ll 1$ where $m_F>few~100$ GeV, so within this scenario, the density of $S$ would overclose the universe.
 Thus, we take $S$ heavy enough to decay into leptons and $F_i$.

 We take the DM candidate to be  the Dirac fermion, $\psi$. The Dirac field can annihilate to lepton and anti-lepton pair via $Y_{R \alpha}$ coupling with a cross section required within thermal freeze-out scenario.
Notice that $\Phi^\prime$ does not appear in the snail diagram. We have added this new scalar doublet to facilitate the annihilation of $\psi \bar \psi$ pair to lepton anti-lepton pairs via
the $Y_{R\alpha}$ coupling. Instead of the  $Y_{R\alpha}$
coupling, we could introduce a coupling of form $Y_{L\alpha} \Phi''e_{R\alpha}^\dagger \psi_L$ where $\Phi''$ is a $SU(2)$ singlet with electric charge equal to that of the electron.
We have taken $Y_{R\alpha}$ coupling instead of $Y_{L\alpha}$ for definiteness. Replacing it with  $Y_{L\alpha}$ does not change the discussion.
Similarly, we could include new colored and charged scalar(s) to introduce Yukawa couplings to quarks and hence annihilation of dark matter pair to quarks. Studying all these possibilities and their potential signature at the LHC is beyond
the scope of the present paper and will be done elsewhere. In summary, in our model DM is composed of $\psi$ along with a subdominant contribution from $\phi_I$.

The following remarks on the $U(1)_{NEW}$ symmetry are in order:
\begin{itemize}
  \item The $U(1)_{NEW}$ not only protects the DM candidate from decay but it also protects the fermions (in particular $\psi$) from having Majorana mass.
  If $\psi$ obtains even a tiny Majorana mass at loop level, it can be decomposed in terms of Majorana mass eigenstates $\psi_1 \equiv (\psi+\psi^c)/\sqrt{2}$
and $\psi_1 \equiv (\psi-\psi^c)/\sqrt{2}$   among which only the lighter one will survive and play the role of the dark matter. With Majorana dark matter,
$\sigma(\psi_1 \psi_1 \to l \bar{l})$ will be either p-wave suppressed or will be suppressed by $m_l^2/m_{\Phi^\prime}^2\ll m_{\psi}^2/m_{\Phi^\prime}^2$ and cannot account
for the observed DM abundance within the thermal freeze-out scenario.
\item Notice that we have assigned opposite $U(1)_{NEW}$   charges to $F_1$ and $F_2$ that appear in the vertices connected to the external $\nu_\alpha$ and $\nu_\beta$ lines.
Without $U(1)_{NEW}$, we could drop $F_2$ and have a lepton number violating mass term of form $F_1^TcF_3$ giving a neutrino mass contribution proportional to $g_\alpha g_\beta$. This will not however help us to make the model more economic because  a mass matrix proportional to $g_\alpha g_\beta$ has only one nonzero mass eigenvalue which cannot account for the realistic neutrino mass structure with at least two nonzero values. To reconstruct the neutrino mass matrix, another field with nonzero coupling component in the direction perpendicular to $g_\alpha$ in the flavor space is required.
\item The $U(1)_{NEW}$ cannot be replaced with a $Z_2$ subgroup of it because $Z_2$ does not forbid Majorana mass for $\psi$. We could however invoke the $Z_3$ subgroup of $U(1)_{NEW}$ under which
$\psi_L \to e^{\pm i 2 \pi /3} \psi_L$ and $\psi_R \to e^{\mp i 2\pi/3} \psi_R$. For neutrino mass generation as well as DM consideration there is no significant difference between these two. The $Z_3$  symmetry allows terms such as $S^3$ but the $U(1)_{NEW}$ symmetry forbids them. The presence of such terms does not change our results. The reason why we have chosen $U(1)_{NEW}$ instead of $Z_3$ is  that $U(1)_{NEW}$ can be eventually gauged to protect against symmetry breaking by quantum gravitational effects. Notice that only new particles are charged  under $U(1)_{NEW}$. The
gauged $U(1)_{NEW}$ can provide a way to have self-interacting DM, which provide a better fit to small scale features.
A kinetic mixing of $U(1)_{NEW}$ with the photon can lead to a direct detection signal.
We will not however try to gauge $U(1)_{NEW}$ here.
\end{itemize}
\section{Lepton Flavor Violating rare decays\label{LFV}}


Before proceeding to discuss contribution to neutrino masses,  dark matter abundance and effects at colliders, let us derive bounds on parameters from searches for LFV rare decays.
The $h_\alpha$ and $g_\alpha$ couplings in Eq. (\ref{Lagrangian}) lead to Lepton Flavor Violating (LFV) rare decays, $l_\alpha \to l_\beta \gamma$ at one loop level. Using formulas in
\cite{Lavoura}, we find that $g_\alpha$ coupling leads to
\begin{equation} \label{LLFFVV} \Gamma(l_\alpha \to l_\beta \gamma)=g_\alpha^2 g_\beta^2\frac{m_\alpha^5}{16 \pi} \frac{[S(t)]^2}{(16 \pi^2)^2 m_S^4} \end{equation}
where \begin{equation} \label{SSS} S(t)= \frac{t-3}{4(t-1)^2}+\frac{\log t}{2(t-1)^3}+\frac{-2t^2+7t-11}{12(t-1)^3}+\frac{\log t}{2(t-1)^4} \end{equation} in which
$t\equiv (m_{F_1^-}/m_S)^2$. $S(t)$ is a monotonously decreasing function with $S(0)=1/6$, $S(1)=1/24$ and $S(\infty ) =1/12t$ so, as expected from decoupling theorem, $\Gamma(l_\alpha \to l_\beta \gamma)$ is suppressed by $1/({\rm Max}(m_S^2,m_{F^-_1}^2))^2$.  The effect of the $h_\alpha$ coupling is given by the same formula replacing $g_\alpha,g_\beta \to h_\alpha,h_\beta$ and $m_{F_1^-}\to  m_{F_2^-}$. If $\Phi^\prime$ couples to more than one flavor,  the $Y_{R \alpha}$ coupling can also lead to similar LFV effects. As mentioned before, to avoid LFV rare decays induced by $Y_{R\alpha}$, we assume $\Phi^\prime$ couples only to one flavor. In the following, we discuss constraints on $g_\alpha$ from LFV bounds.

The best present bounds on LFV rare decay branching ratios are ~\cite{pdg}
\be \label{mue} Br(\mu \to e \gamma)<5.7\times 10^{-13}~,\ee
\be \label{taue} Br(\tau \to e \gamma)<3.3\times 10^{-8} \ee
and
\be \label{taumu} Br(\tau \to \mu \gamma)<4.4\times 10^{-8}~.\ee

From Eq. (\ref{mue}), we find
\be \label{gegmu} g_eg_\mu \stackrel{<}{\sim} 10^{-3} \frac{{\rm Max}(m_S^2,m_{F^-_1}^2)}{{\rm TeV}^2} \ee
and from Eqs. (\ref{taue},\ref{taumu}), we find
\be \label{gegmugtau} g_eg_\tau, g_\mu g_\tau \stackrel{<}{\sim}  \frac{{\rm Max}(m_S^2,m_{F^-_1}^2)}{{\rm TeV}^2} .\ee
Similar consideration and bound hold valid for the $h_\alpha$ coupling, replacing $m_{F_1^-}\to  m_{F_2^-}$.

\section{Neutrino masses\label{NuM}}
\begin{figure}[t!]
 \centering
  \includegraphics[scale=1.3]{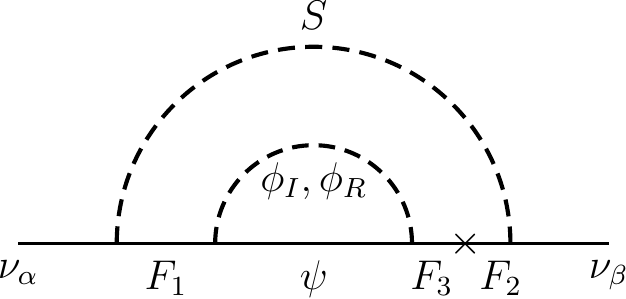}
  \caption{\it Diagram giving mass to neutrinos. $``\times " $ indicates the $m_M$ mass term insertion which violates lepton number conservation.
    \label{rainbow}}
\end{figure}
For simplicity, let us set $Y_1^\prime=Y_2^\prime=0$.  Discussion for nonzero $Y_1^\prime$ and $Y_2^\prime$ will be
similar. In this model, we  have only one diagram contributing to neutrino mass. That is of form of snail diagram shown in Fig (2-d), where $S_1$, $S_2$ and $F_4$ should be respectively identified with $S$, $\Phi^0$ and $\psi$ of our model. Instead of using $\lambda \langle H\rangle^2 (\Phi^0)^2$ mass insertion approximation,
  we can have mass eigenstates $\phi_I$ and $\phi_R$ (imaginary and real components of $\Phi^0$) propagating in the inner loop as shown in Fig. \ref{rainbow}. Going to mass basis $\phi_R$ and $\phi_I$, the contribution of these fields propagating in the inner loop will be respectively given by factors $(Y_1/\sqrt{2})(Y_2/\sqrt{2})[1/(p^2-m_R^2)]$ and  $(iY_1/\sqrt{2})(iY_2/\sqrt{2})[1/(p^2-m_I^2)]$ so the sum of two contributions will be proportional to
$$\frac{Y_1Y_2(m_R^2-m_I^2)}{2(p^2-m_I^2)(p^2-m_R^2)}.$$
We use mass insertion approximation for $\langle F_3 F_2^T \rangle$ propagator: $k^2 m_M/[(k^2-m_{F_3}^2) (k^2-m_{F_2}^2)]$. Putting all these together we find that the two-loop snail diagram contribution to neutrino mass is given by
$$ (m_{\nu})_{\alpha \beta}= (g_\alpha h_\beta +g_\beta h_\alpha) m_M \frac{Y_1 Y_2}{2} (m_R^2-m_I^2)\int \frac{d^4p}{(2\pi)^4} \int \frac{d^4 k}{(2\pi)^4} $$
$$ \frac{1}{k^2-m_S^2} \frac{k\cdot \sigma}{k^2-m_{F_1}^2} \frac{( p+k) \cdot \bar{\sigma}}{(k+p)^2-m_{\psi}^2}\frac{1}{(p^2-m_R^2)(p^2-m_I^2)} \frac{k^2 }{(k^2-m_{F_2}^2)(k^2-m_{F_3}^2)} \ .$$
Without loss of generality, we can go to a basis where $g_\alpha$ takes the form of $(0,0,g)$. We still have the freedom to rotate $h_\alpha$ in the direction $(0,h_1,h_2)$. In this basis, the first row and column of $m_\nu$ vanishes so with this field content one of neutrino mass eigenvalues will be zero. The mass scheme will be therefore hierarchical but the mixing parameters and CP-phases can be reconstructed with proper choice of $g_\alpha$ and $h_\beta$. To obtain non-hierarchical scheme, we can add another singlet $S$ coupled to $L$.
Using Feynman parameters we find
$$
 (m_{\nu})_{\alpha \beta}= \frac{(g_\alpha h_\beta +g_\beta h_\alpha)}{16} m_M Y_1 Y_2\frac{ (m_R^2-m_I^2)}{(16\pi^2)^2}
 I(m_{F_1},m_{F_2},m_{F_3},m_S,m_\psi,m_I,m_R)$$
 where $ I(m_{F_1},m_{F_2},m_{F_3},m_S,m_\psi,m_I,m_R)$ is defined as $$
 \int_0^1dy\int_0^{1-y} {dx}\int_0^1 da_1 \int_0^{1-a_1} da_2 \int_0^{1-a_1-a_2} da_3\int_0^{1-a_1-a_2-a_3} da_4 ~\frac{1-x}{A}$$ in which $A$ is equal to
$$(a_1 m_{F_1}^2+a_2 m_{F_2}^2+a_3 m_{F_3}^2 +a_4 m_{S}^2)x(1-x)+(1-a_1-a_2-a_3-a_4)(xm_{\psi}^2+ym_I^2+(1-x-y)m_R^2).$$
Notice that $A$ is a positive definitive quantity over the whole integration range. Thus, the integration $ I$ is a finite quantity as expected.
$\psi$ is the lightest field propagating in the loops. Let us denote the mass of the  heaviest field propagating in the loop by $m_{max}$. We can then write $ I(m_{F_1},m_{F_2},m_{F_3},m_S,m_\psi,m_I,m_R)=b/m_{max}^2$
where $b$ is a number. For $m_\psi/m_{max}$ (and therefore the rest of ratios) varying  between $\sim 0.1$ to 1, the value of $b$ varies   in the range $O(0.01)$-$O(0.1)$.
 The neutrino mass can be then estimated as
\be \label{mnu-estimate} m_\nu \sim (0.01-0.1~{\rm eV}) Y_1 Y_2 \frac{g\times h}{10^{-1}\times 10^{-2}} \frac{m_M}{5 ~{\rm GeV}}\frac{(m_R^2-m_I^2)/m_{max}^2}{1/20}.\ee
Notice that $m_R^2-m_I^2\sim \lambda \langle H^0\rangle^2$. Taking $\lambda \sim 0.5$ and $m_{NEW} \sim ({\rm few ~TeV})$, it seems to be natural to have
$(m_R^2-m_I^2)/m_{R}^2\leq (m_R^2-m_I^2)/m_{NEW}^2 \stackrel{<}{\sim} 0.1$. As we will discuss in sec. \ref{annihilation}, $(m_R-m_I)/m_R$ should be smaller than $\sim 0.05$ to facilitate the coannihilation of
$\phi_I$ and $\phi_R$  ({\it e.g.,} $\phi_I \phi_R \to Z^* \to SM$) in the early universe and hence prevent over-closure of the universe by lighter component of $\phi_I$ and $\phi_R$.

The following points are in order:
\begin{itemize}
\item To make the estimate in Eq. (\ref{mnu-estimate}), we have taken $g_\alpha h_\beta\sim 10^{-3}$. As we saw see in section \ref{LFV}, for $m_{NEW}\sim 1$~TeV, the upper bounds on ${g_e g_\mu}$ and  ${h_e h_\mu}$
from Br($\mu \to e \gamma$) are of order of $10^{-3}$ so we expect an observable effect in near future at searches for $\mu \to e \gamma$.
Within this model, saturating bounds on Br($\tau \to \mu \gamma$) or  Br($\tau \to e \gamma$) can be possible only if $g_\alpha\sim 10^{-3} \ll h_\alpha\sim 1$ or
 $g_\alpha\sim 10^{-3} \gg h_\alpha\sim 1$.
\item To arrive at Eq. (\ref{mnu-estimate}), we have used mass insertion approximation for the treatment of mass term mixing $F_2$ and $F_3$, $m_M$. Taking $m_M=5$ GeV and $m_{F_i}\sim $TeV, this approximation is valid. Taking smaller $m_M$ requires $Y_1,Y_2 \gg 1$ which leads to non-perturbativity.
    \item In the range $m_{\Phi}\sim m_F \sim m_{max}\sim 1~{\rm TeV}-100~{\rm TeV}$ and $m_M\sim 5~{\rm GeV} (m_{max}^2/{\rm TeV}^2)$, we obtain desired values of $m_\nu$ satisfying bounds from LFV as well as collider searches and we still remain in the perturbativity range: $Y_1,Y_2<1$ and $m_M\ll m_F$.
The lower part of this range can be probed at second phase of the LHC, but the range $m_F,m_\phi>10$ TeV is out of the reach of the LHC.
\end{itemize}
\section{Annihilation to lepton pair \label{annihilation}}
As discussed in section \ref{sec:snail-model-DM}, we choose the main dark matter component to be $\psi$ which annihilates to a pair of leptons.
The annihilation cross section to a  charged lepton pair of flavor $\alpha$ can be written
as
\be \label{annihi} \langle \sigma( \psi \bar \psi \to \ell_\alpha \bar\ell_\alpha) v \rangle = \frac{|Y_{R\alpha}|^4 }{32 \pi} \frac{ m_{\psi}^2}{
(m_{\psi}^2+(m_{\phi^{\prime -}})^2)^2}.
\ee
A similar equation can be rewritten for annihilation to a $\nu_\alpha \bar\nu_\alpha$ pair by replacing $m_{\phi^{\prime -}}$ with $m_{\phi^{\prime 0}}$.
To avoid large LFV effects, we assume that only one  flavor component of $Y_{R\alpha}$ is nonzero.
Taking $\langle \sigma_{tot} v\rangle = 3  \times 10^{-26} ~{\rm cm}^3 {\rm sec}^{-1}$ (as predicted within the thermal freeze-out scenario) and typical values $m_\psi=300$ GeV and $m_{\phi^{\prime -}}=m_{\phi^{\prime 0}}=400$ GeV
we find $Y_R=0.55$. In general, we obtain
\be \label{YR}
m_{\phi^{\prime -}},~m_{\phi^{\prime 0}} \leq 1.4 Y_{R\alpha}^2 {\rm TeV} \ee
where equality corresponds to the limiting case of $m_\psi \to m_{\phi^{\prime -}}\simeq m_{\phi^{\prime 0}}$.

The large $Y_{R\alpha}$ coupling   will not however affect the  lepton or heavy meson decays because they are not heavy enough to emit $\psi$.
This large coupling can cause dips in the spectrum of very high energy  cosmic neutrinos at ICECUBE due to scattering off the  DM distributed all over the universe. The resonance energy is at
$E_{res} \sim (m_{\phi^{\prime 0}})^2/m_{\psi}\sim {\rm few} ~100~ {\rm GeV} $.  For a given $m_{\phi^\prime}$, decreasing $m_\psi$, the value of $E_{res}$ and as a result the position of the dip shifts towards higher energies. One should however bear in mind that  by decreasing $m_{\psi}/m_{\phi^\prime}$ the required $Y_R$  increases and eventually enters non-perturbative regime.

Data from the region close to galaxy center from Fermi-LAT shows a hint of GeV range gamma excess. One of the explanations is the annihilation of 10 GeV DM pairs to lepton pairs \cite{Celine}. It is tantalizing to try to accommodate this signal within our model.
Now, following Ref \cite{Lacroix:2014eea}, if we set $\langle \sigma( \psi \bar \psi \to l \bar l) v \rangle =0.86 \times 10^{-26} ~{\rm cm}^3 {\rm sec}^{-1}$ and
$m_{\psi}\sim 10 $ GeV, we obtain
$$Y_{R }=0.5 (m_{\phi^{\prime -}}/100~{\rm GeV}) (10 ~{\rm GeV}/m_{\psi})^{1/2}\ . $$
Notice that we have taken $\phi^\prime$ to be relatively light.
From the first run of the LHC there, there is already a lower bound of  325 GeV on the  mass of new charged scalar such as $\phi^{\prime -}$ whose decay lead to the electron or the muon plus missing energy \cite{Aad}.  Bounds on such scalar coupled to only tau is weaker: $m_{\phi^{\prime -}}>90$ GeV \cite{stau}. As a result, for annihilation to tau pair, the value of $m_{\phi^{\prime -}}$ satisfies the present bound. For heavier values of $\phi^{\prime -}$, we eventually enter non-perturbative regime. A more recent analysis of the gamma ray excess finds a better fit with $m_\psi \sim 50$ GeV and $\langle \sigma(\psi \bar\psi \to b \bar b)\rangle \sim 10^{-26}~ {\rm cm}^3 {\rm sec}^{-1}$ \cite{McCabe}. This can be achieved with a coupling of form $Y_b \bar{b}_R\psi \phi^{''}$ where $\phi^{''}$ is a colored and charged scalar singlet under $SU(2)$. From the LHC bounds, this scalar should be heavier than 620~GeV \cite{sb}. The annihilation cross section of $\psi \bar{\psi} \to b \bar{b}$ is given by Eq. (\ref{annihi}) replacing $\phi^\prime$ with $\phi^{''}$ and multiplying by a factor of three to account for the color degrees of freedom. To accommodate the signal with $m_\psi$ and $m_{\phi^{''}}\sim 700$~GeV, $Y_b$ should be of order of one. One should however bear in mind that DM origin of gamma ray excess has been questioned in a series of publication \cite{Gabi}.

As discussed before the lightest neutral component of $\phi$ ({\it i.e.,} $\phi_I$) can be another DM component if it is lighter than $\phi^-$. For $|m_R-m_I|\stackrel{<}{\sim} m_R/20$,  coannihilation via
$\phi_I\phi_R \to Z^*\to SM$ will render its abundance negligible.

\section{Signature at the LHC\label{LHC}}
In this model, there are several fields with electroweak interactions that can be pair produced at the LHC provided that they are light enough. As discussed  in sec. \ref{NuM}, $\Phi$
and $F_i$ fields propagating in the loops that contribute to $m_\nu$ can have masses in the range 1~TeV-100~TeV.  As discussed in sect. \ref{sec:snail-model-DM}, we take $\Phi$ to be lighter than $F_1$ and $F_3$. As result, via large $Y_1$ and $Y_2$
couplings, the components of $F_2$ and $F_3$ will decay as $F_i^- \to \psi \phi^-$ and $F_i^0 \to \psi \phi^0_{I(R)}$. The $\psi$ particle as well as  $\phi_I$  will appear as missing energy.
Via tree-level $Z^*$ exchange,
$\phi_R \to \phi_I \nu \bar{\nu}, ~\phi_I l \bar{l}$.

While $\Phi$ and $F_i$ particles can be too heavy to be produced at the LHC, as we saw in sect. \ref{annihilation}, there  is an upper bound on the masses of the $\Phi^\prime$
components. Thus, if this model is realized in nature, it is guaranteed that the components of $\Phi^\prime$ will be pair produced at the second run of the LHC, leading to
the following signals:
\begin{itemize} \item Mono-lepton plus missing energy signal through
$u \bar{d} \to \phi^{\prime +} \phi^{\prime 0} \to (l^+\psi) (\nu \bar{\psi})$
and the charge conjugate processes.
\item Two-lepton plus missing energy signal through
$ u\bar{u},d \bar{d} \to \phi^{\prime +} \phi^{\prime -} \to (l^+\psi) (l^- \bar{\psi}).$
\item Missing energy through
$ u\bar{u},d \bar{d} \to \phi^{\prime 0} \bar{\phi}^{\prime 0} \to (\bar{\nu}\psi) (\nu \bar{\psi}).$
\end{itemize}
As discussed in section \ref{annihilation}, the present lower bounds on the masses of scalars whose decay lead to missing energy plus muon and electron \cite{Aad} and tau lepton \cite{stau} 
are respectively 325~GeV and 90~GeV. 
In fact, phenomenology of $\Phi^\prime$ doublet at the LHC (both  production mechanism as well as signature of  the decay product) is very similar to that of left-handed slepton doublet in the framework of Minimal Supersymmetric Standard Model (MSSM).
As mentioned before, we assume $\phi^\prime$ to couple mainly to only one flavor to avoid LFV rare processes. If this flavor happens to be the second generation, the signals at the LHC will
be cleaner.
In this case, we expect a contribution to $(g-2)_\mu$ which we elaborate on in the next section.
\section{Muon magnetic dipole moment}
In this model, there are several particles that couple to the muon and can give rise to $(g-2)_\mu$ at one loop level. Considering the bounds in Eq. (\ref{YR}) on the mass and coupling
of $\phi^\prime$, it can give largest contribution to $(g-2)_\mu$ if the $Y_{R\alpha}$ coupling is to the muon flavor.
The $Y_{R\mu}$ coupling leads to
$$\delta \frac{g-2}{2}= \frac{Y_{R\mu}^2}{16 \pi^2} \frac{m_\mu^2}{m_{\phi^{\prime -}}^2} K(r)$$
where
$$ K(r)=\frac{2 r^2+5r-1}{12(r-1)^3}-\frac{r^2 \log r}{2(r-1)^4}$$
in which $r=(m_\psi^2/m_{\phi^{\prime -}}^2)$.
Taking  $m_{\phi^\prime}\sim 100~{\rm GeV}-1~{\rm TeV}$ and   $Y_{R\alpha}\sim 1$ (see Eq. \ref{YR}), we find that
$(g-2)_\mu/2\sim 10^{-11}-10^{-12}$ which is well
below the current sensitivity limit \cite{pdg}.
\section{Conclusions}
\label{sec:conclusions}
Following the ``recipes'' developed in \cite{Farzan:2012ev}, we have built a model in which neutrinos receive Majorana mass via a two-loop diagram with topology
of ``snail diagram'' depicted in Fig. \ref{fig:crab-plus-snail}-d and in Fig. \ref{rainbow}. The particles propagating in the loops are new scalars and fermions charged under $SU(2)\times U(1)$. The field content is given in table \ref{tabl}. The lepton number is explicitly broken by fermion mass $m_M$ (see Eq. \ref{mM}) so the neutrino masses are proportional to $m_M$ as seen in Eq. (\ref{mnu-estimate}). Following the argument in Ref. \cite{Farzan:2012ev}, we confirm that the two-loop snail diagram is the leading contribution to neutrino mass. Within this model the neutrino mass scheme is predicted to be hierarchical with one vanishing mass eigenvalue. The model respects a global $U(1)_{NEW}\times Z_2$ symmetry which stabilizes two of new particles: $\phi_I$, the imaginary part of the neutral component of $\Phi$ and $\psi$, a singlet under electroweak group. We assume the mass splitting between $\phi_I$ and $\phi_R$ (the real component of $\phi^0$) is small enough to allow efficient co-annihilation in the early universe. $\phi_I$ is therefore only a sub-dominant component of dark matter. This assumption turns out to be natural within our model and does not need any fine-tuning.

The dominant component of dark matter is Dirac fermions $\psi$ that can annihilate to
a pair of standard model fermions via a Yukawa coupling  involving  new scalar $\Phi^\prime$.
 In order to obtain the observed abundance of dark matter within freeze-out scenario ({\it i.e.,} $\langle \sigma(\psi \bar\psi \to f \bar
f)v \rangle \sim 1$~pb), the mass of ${\Phi^\prime}$ should be less than 1.5 TeV (see Eq. (\ref{YR})). This means the components of $\Phi^\prime$ can be eventually produced at the LHC via electroweak interactions and discovered through their signature of decay to standard model fermions plus missing energy. Moreover the corresponding Yukawa coupling should be of order of one. To avoid LFV rare decay, we assume $\Phi^\prime$ couples only to one flavor. If this flavor is the muon, the discovery potential of the LHC will be higher. The contribution to $(g-2)_\mu$ is then predicted to be one or two orders of magnitude below the present sensitivity. The coupling of the scalar singlet, $S$ to leptons ({\it i.e.,} $g_\alpha$ and $h_\alpha$) should involve more than one flavor to reconstruct the neutrino mass matrix structure. This in turn leads to LFV rare decays. From values of neutrino mass, we expect the $\mu \to e \gamma$ signal to be  around the corner.
\section{Acknowledgments}
The author would like to specially thank Diego Aristizabal Sierra with  whose collaboration  the early stages of this work was done. She also thanks his daughter  Sofia
Aristizabal for suggesting names for the ``snail'' and ``crab'' diagrams. The author also appreciates Majid Hashemi for useful remarks and discussions on possible signature at the LHC. She  would like to thank NORDITA where this project started  as well as Liege Univ where a part of this work was done. She
acknowledges partial support from the European Union FP7 ITN
INVISIBLES (Marie Curie Actions, PITN- GA-2011- 289442).

\end{document}